\documentclass[sigconf]{acmart}

\usepackage{booktabs} 
\usepackage[utf8x]{inputenc} 
\usepackage{balance}
\usepackage{titlesec}
\usepackage{multirow}
\usepackage{caption}
\usepackage{graphicx}

\copyrightyear{2019}
\acmYear{2019}
\acmConference[CUI 2019]{1st International Conference on Conversational User Interfaces}{August 22--23, 2019}{Dublin, Ireland}
\acmBooktitle{1st International Conference on Conversational User Interfaces (CUI 2019), August 22--23, 2019, Dublin, Ireland}
\acmPrice{15.00}
\acmDOI{10.1145/3342775.3342786}
\acmISBN{978-1-4503-7187-2/19/08}

\title[What's in an accent?] {What’s in an accent? The impact of accented synthetic speech on lexical choice in human-machine dialogue}

\begin{document}

\author{Benjamin R. Cowan}
\email{benjamin.cowan@ucd.ie}
\affiliation{%
  \institution{University College Dublin}
  \city{Dublin}
  \country{Ireland}
}

\author{Philip Doyle, Justin Edwards, Diego Garaialde}
\email{philip.doyle1, justin.edwards, diego.garaialde@ucdconnect.ie}
\affiliation{%
  \institution{University College Dublin}
  \city{Dublin}
  \country{Ireland}
}


\author{Ali Hayes-Brady}
\email{ali.hayes-brady@monash.edu}
\affiliation{%
  \institution{Monash University}
  \city{Melbourne}
  \country{Australia}
}
\author{Holly P. Branigan}
\email{Holly.Branigan@ed.ac.uk}
\affiliation{%
  \institution{University of Edinburgh}
  \city{Edinburgh}
  \country{UK}
}
\author{João Cabral}
\email{cabralj@scss.tcd.ie}
\affiliation{%
\institution{Trinity College Dublin}
  \city{Dublin}
  \country{Ireland}
}

\author{Leigh Clark}
\email{leigh.clark@ucd.ie}
\affiliation{%
  \institution{University College Dublin}
  \city{Dublin}
  \country{Ireland}
}

\renewcommand{\shortauthors}{B. R. Cowan et al.}

\begin{abstract}
The assumptions we make about a dialogue partner's knowledge and communicative ability (i.e. our partner models) can influence our language choices. Although similar processes may operate in human-machine dialogue, the role of design in shaping these models, and their subsequent effects on interaction are not clearly understood. Focusing on synthesis design, we conduct a referential communication experiment to identify the impact of accented speech on lexical choice. In particular, we focus on whether accented speech may encourage the use of lexical alternatives that are relevant to a partner’s accent, and how this is may vary when in dialogue with a human or machine. We find that people are more likely to use American English terms when speaking with a US accented partner than an Irish accented partner in both human and machine conditions. This lends support to the proposal that synthesis design can influence partner perception of lexical knowledge, which in turn guide user's lexical choices. We discuss the findings with relation to the nature and dynamics of partner models in human machine dialogue. 

\end{abstract}

%
%



\maketitle

\section{INTRODUCTION}
Psycholinguistic research suggests that language choices in human-human dialogue (HHD) are influenced by the assumptions we make about our partners as communicative and social beings (i.e. our partner models) \cite{branigan2011role,cowan2015does}. People tend to estimate a conversation partner's knowledge and communicative abilities, and form their utterances accordingly. This potentially complex process is simplified by using a range of cues as partner model heuristics. For instance, we might use accent and/or social cues \cite{clark1996using,nickerson1999we} alongside beliefs about the social distribution of knowledge in particular social groupings (e.g. opticians or whisky enthusiasts) \cite{Fussell1992} to make inferences about what a partner might know and understand. 

Such perspective taking does not only occur in HHD. Current literature supposes that many language choices in human-machine dialogue (HMD) are adaptive and driven by users’ partner models of a machine as a dialogue partner \cite{amalberti1993user,brennan1998grounding,LEBIGOT2007983,meddeb2010didn,branigan2011role}. Compared to HHD, users tend to use simpler grammatical structures, use more words in their descriptions, fewer pronominal anaphors (e.g. her/him; he/she) and make simpler and more similar lexical choices when in HMD \cite{amalberti1993user,KENNEDY198837}. Although little is currently known about what constitutes and drives partner models in an HMD context, design decisions around issues like the humanness of speech synthesis or the language a system produces have been proposed as influential \cite{cowan2015voice,luger2016like,cowan2017can}.

With speech widely used as a primary interface modality in a number of applications and devices (e.g. Siri, Google Assistant, Amazon Echo) it is imperative that we develop an understanding of the mechanisms that drive language choices in HMD interactions \cite{clark2018state}. Here we explore how a dimension of humanness, namely accent, may act as a cue to partner modeling and how this in turn may impact language choices in HMD. The research presented focuses on the use of synthesised Irish or US accents. It investigates how this impacts the use of Hiberno or American English lexical items when naming objects that have Hiberno and American English lexical alternatives. Through using a referential communication task we found that in both HHD and HMD people were more likely to use American-English terms when interacting with US accented partners than when interacting with Irish accented partners. This supports the proposition that partner model assumptions of lexical knowledge and understanding are influenced by partner accent, informing lexical choices when such lexical alternatives are possible. 

\section{RELATED WORK}

\subsection{Perspective-Taking in Dialogue}
Imagine that someone asks for directions. How do we decide what language to use so as to most appropriately communicate these? Research into HHD highlights that verbal and nonverbal cues (such as speaker nationality, language proficiency, profession) are important in helping to assess the characteristics of our interaction partners \cite{nickerson1999we}. These characteristics are then used to develop a initial (or global) partner model \cite{brennan2010two}, informing what language we decide to use \cite{Fussell1992}. The initial directions we give are then designed specifically with this model in mind, yet this model can be dynamically altered as other cues are processed and negotiated in the dialogue. These inferences allow us to develop an impression of common ground; mutual knowledge, assumptions and beliefs shared between interlocutors when they converse, thought to play an important role in communication efficiency and success \cite{bromme2001expertise,clark1996using}. 

There is strong evidence that we adjust our language based on our assumed knowledge of our audience (termed audience design \cite{bell1984language}). When people are asked to describe items for their friends, they adapt their descriptions to their friend’s knowledge, and these adjustments lead to higher accuracy in identification \cite{fussell1989understanding}. Shared context and shared experiences with interlocutors, as well as social category cues such as accent, attire and context have all been highlighted as important to partner modeling \cite{nickerson1999we}.

Much of the work in HMD suggests that partner modeling is a major determinant of language choices (e.g. \cite{branigan2011role,EDLUND2008630,oviatt1998linguistic}). People tend to see automated partners as poorer interlocutors and alter their language choices and speech behaviours with this in mind \cite{branigan2011role,oviatt1998linguistic}. For example, people are more likely to converge (or align) with their partner’s choice of referring expression when they believe their partner to be a computer rather than a human, so as to ensure they are understood. In addition, they align more when they are led to believe that the computer partner is a ‘basic’ interlocutor with restricted capability than a partner with more advanced capability \cite{branigan2011role}. A study looking at people's speech in telephone conversation concerning air-fares and timetables found that people’s linguistic behaviour changed depending on whether they believed their partner to be a human or a computer \cite{amalberti1993user}. Other work has echoed these findings \cite{bell1999interaction,KENNEDY198837}. Although there is an assumption that people’s beliefs of partner abilities affect language choice in this context, it is still not clear what factors drive these beliefs in interaction. Recent work has proposed that, although we generally see machines as limited communicative partners \cite{clark2019makes,branigan2011role}, design decisions used to portray humanness of speech systems may be important drivers of people’s initial partner models \cite{luger2016like,cowan2017can}, acting as a metaphor to support interaction. Crucially, these design decisions can overinflate perceptions of system competence \cite{moore2017spoken,luger2016like}, significantly affecting the quality of interaction. 

\subsection{Accent as a cue to partner modeling}
Accent is an important characteristic to consider when developing human-like synthesis for speech interfaces, and may have a significant impact on partner models. In HHD, accents signpost a speaker’s social identity and socio-linguistic background \cite{Ikeno2007}. This can strongly influence perceptions of a speaker, eliciting specific stereotypes and assumptions associated with a particular accent \cite{ryan_integrative_1982}. Likewise in HMD, accent and other indicators of a machine’s national origin, conveyed through speech, play a strong role in user behaviour and perceptions. \citet{dahlback2001spoken} looked at self-disclosure behaviours when people were interviewed by a virtual agent with an accent denoting they were either of the same, or, of a different nationality to the user. In the study US participants exhibited strong preferences for the US-accented virtual agent. They spoke more to the US-accented agent during dialogue, found it more sociable and felt they gave more honest answers than when interacting with a Swedish-accented interviewer. Swedish participants also reported the same feelings toward the Swedish-accented agent \cite{dahlback2001spoken}. Elsewhere, similar work looking at voice-based judgments people make about partners \cite{dahlback2007similarity} asked Swedish and US participants to listen to tourist information about New York and Stockholm. This information was relayed through synthesised speech with either a US or Swedish accent. Participants rated the information as more valuable and likeable when delivered in an accent that matched their own, irrespective of which city was being described \cite{dahlback2007similarity}. The authors suggest these findings support the idea of a similarity-attraction effect, whereby we judge communicative systems that are similar to ourselves (e.g. in accent, gender and/or personality) more positively \cite{dahlback2007similarity}, irrespective of their perceived expertise of the context in question.

Research on robotic agents has also demonstrated that participants’ judgments about an agent’s abilities are influenced by both the perceived nationality of the agent and the content that it is being asked to process \cite{lee2005human}. Participants were asked to judge the likelihood that a robot would be able to know and recognize a set of landmarks from New York and Hong Kong. Half of the participants were told that the robot was built in New York and were shown a video of it interacting with experimenters in English. The other half were told that it was built in Hong Kong with a video of it interacting with the experimenters in Cantonese. They found that people used these cues in HMD similarly to how they used them in HHD, whereby people perceived landmarks familiar to each nationality as more identifiable to those partners \cite{lee2005human}. 
\subsection{Research Aims}
Accent and partner nationality perceptions seem to affect user perceptions in speech and language interaction with machines. Yet it is not clear whether this might also affect user language production. This may be particularly relevant in situations where participants need to describe or name objects where lexical alternatives relevant to the accent exist (e.g., nappy vs diaper for Hiberno and American-English respectively). This work looks to explore whether partner nationality, in particular whether the partner uses an Irish or US accent, has a significant impact on object naming in HMD. We hypothesise that partner nationality will have a statistically significant effect on the number of  American English terms used to refer to objects. Specifically we predict that more American-English (and less Hiberno-English) terms will be used to refer to objects when interacting with a US accented partner than when interacting with an Irish accented partner (H1). We also hypothesise that, due to machine partners being seen as less flexible than human partners, there will be a statistically significant interaction between partner type and perceived partner nationality (H2), whereby there will be a stronger effect of accent on lexical choice in the machine compared to human partner conditions.

\begin{table*}[t]
\caption{Response rates for target item lexical alternatives}
\label{my-label}
\begin{tabular}{@{}cccc@{}}
\toprule
\textbf{Item number} & \textbf{\begin{tabular}[c]{@{}c@{}}American-English Names\\ (number of responses)\end{tabular}} & \textbf{\begin{tabular}[c]{@{}c@{}}Hiberno-English Names\\ (number of responses)\end{tabular}} & \textbf{Other} \\ \midrule
1* & diaper (1) & nappy (31) &  2 \\
2* & wrench (15) & spanner (18) & 1 \\
3 & candy (1) & sweet (17) & 16 \\
4* & eggplant (6) & aubergine (23) & 5 \\
5 & zucchini (2) & courgette & 8 \\
6* & broiler (1) & grill (32) & 1 \\
7* & elevator (11) & lift (21) & 2 \\
8 & attorney (0) & solicitor (10) & 24 \\
9 & trashcan (0) & bin (31) & 3 \\
10 & transmission (0) & gearbox (12) & 22 \\
11 & oatmeal (1) & porridge (20) & 13 \\
12* & flashlight (3) & torch (31) & 0 \\
13 & sweater (0) & jumper (34) & 0 \\
14 & drugstore (0) & pharmacy (34) & 0 \\
15 & intersection (0) & crossroads (18) & 16 \\
16 & bangs (0) & fringe (29) & 5 \\
17 & freeway (0) & motorway (24) & 10 \\
18* & ladybug (3) & ladybird (31) & 0 \\ \bottomrule
\end{tabular}
\end{table*}

\section{METHOD}
\subsection{Participants}
Thirty-four participants (16 female, 18 male; M age=28.69yrs, SD=\\13.35yrs) recruited from a European university took part in the study. All were native or near native Hiberno-English speakers with 31 being Irish and 3 being British nationals. Participants were recruited from staff and students at the university and came from a variety of disciplines including chemistry, history, psychology and information science.  Twenty-nine reported previous use of speech interface technologies, with most reporting infrequent use (7 point Likert scale: 1=Very Infrequently- 7=Very Frequently; M=2.83; SD=1.77). Participants were given a €10 voucher as an honorarium for participating in the research.   

\begin{figure}
  \includegraphics[scale=0.15]{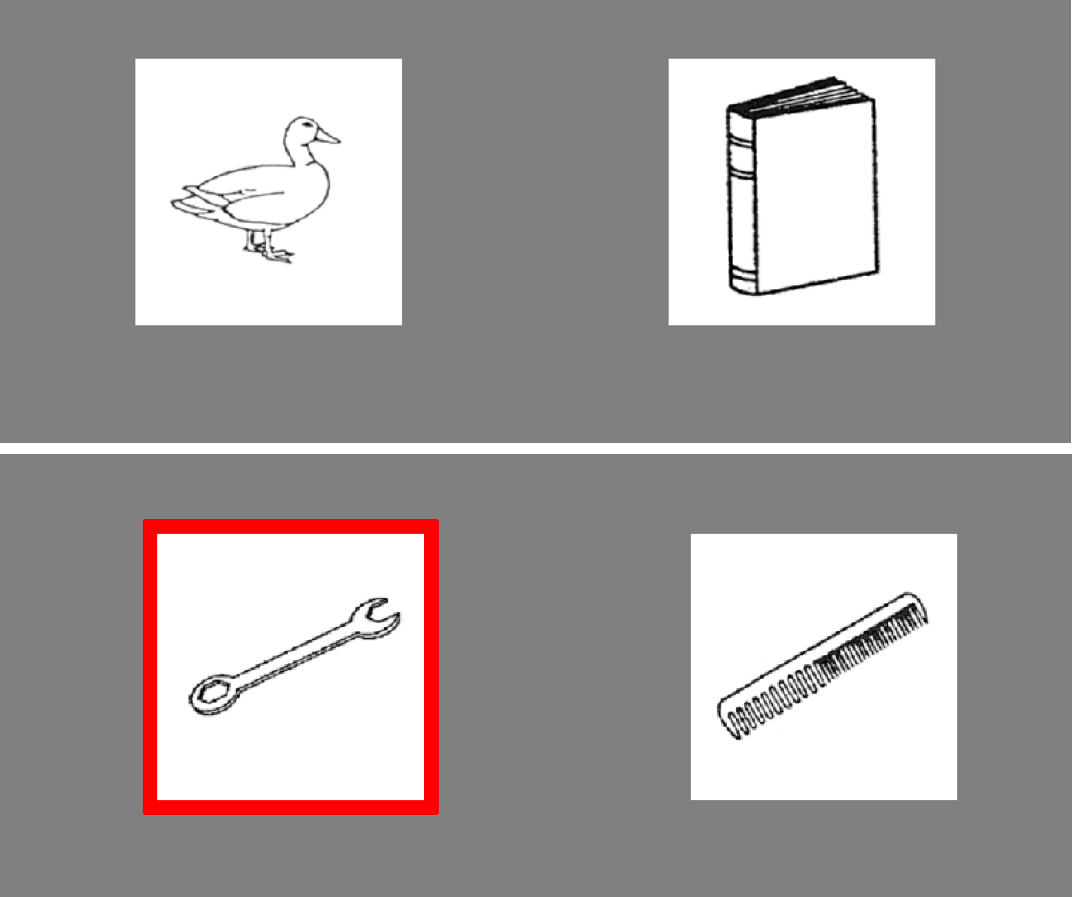}
  \caption{Example screenshots of matching turn (a-Top) and naming turn (b-Bottom) for communication game.}
  \label{fig:fig2}
\end{figure}

\subsection{Experiment Task}
\subsubsection{Referential Communication Task- Picture Naming}
Participants were asked to complete a referential communication task similar to those previously used to research lexical choice in HHD and HMD (e.g. \cite{branigan2011role,cowan2015does}). The task took the form of a picture naming and matching game, which participants played with a partner. In the game, the participant and the partner took turns to either name an object depicted in an image on screen (naming turn- see Figure 1b) or select an image that matched their partners description from an array of two images (matching turn- see Figure 1a). The game was built and run using PsychoPy \cite{peirce_building_2018} v. 1.85.6.

\subsubsection{Target Item Development \& Selection}
The game included 18 target items, that appeared in the naming turns in the game. The objects depicted in these, varied in the words that could be used to name the object in either Hiberno-English or American-English (e.g. Lift or Elevator respectively). As there were no resources available offering Hiberno vs. American-English lexical comparisons directly, an initial selection of 40 possible word pairings of British-English vs. American-English terms were collated using the Oxford English Dictionary and other online lists. Potential pairings were then screened in a separate study to 1) ensure that British-English terms identified were rated as likely to be understood by Hiberno-English speakers and 2) to identify the most effective word pairs to include in the experiment. Here twenty-four Hiberno-English speaking participants rated how likely it would be for another Irish person and an American person to understand each of the words chosen. The rating scale ranged from 1 to 7: 1 = 'Very likely to be understood by an Irish person but not an American person'; 7 = 'Very likely to be understood by an American person but not an Irish person'; a mid-rating score of 4 indicated a rating that the term was equally likely to be understood by both an Irish and an American person. Participants were also given the option of stating that they thought the word would be understood by Neither. Word pairs were then chosen to be included in the experiment based on those that had the largest difference between their mean and mode ratings on this scale. Another consideration was whether it was possible to clearly draw the item so it could elicit the lexical alternatives desired. For instance, word pairings such as freshman and undergraduate attained large mean and mode rating differences (which suggested they clearly differed in how likely people felt those words to be recognisable to an American and Irish person) but were deemed difficult to clearly depict in an image so as to elicit the lexical alternatives consistently. 

From this a final list of 18 word pairings was selected. As the list was derived from terms perceived by an Irish cohort to be most recognisable to Irish or American people, we regard them from here on in as Hiberno-English and American-English respectively. Pictures were then generated depicting the item relevant for these word pairings, ensuring to use a similar style to those developed by \cite{snodgrass1980standardized}, and used in previous HMD referential communication games (e.g. \cite{branigan2011role,cowan2015does}). A list of the final target items and the predicted Irish and American names are included in Table 1. The images generated for these items are included in supplementary material. 

\subsubsection{Filler Items}
The game also included 75 filler naming turns. These were included to mask the focus of the study being on the experiment target items. The items depicted in the filler images did not vary in potential American or Hiberno-English lexical alternatives. 

\subsection{Game Structure}
The game consisted of a total of 93 naming-matching turns. Within the naming turns, 18 target items and 75 filler items were displayed. As depicted in Figure 1b, two images were displayed in the naming turns (either a target item or a filler item and a distractor image). After naming the object in the image, participants pressed enter to move to the matching turn. In the matching turn two filler images were displayed, with the participant hearing a description from their partner. Participants had to press either “1” or “2” to select which image matched the description given by their partner. No target images were displayed in the matching turns so as to ensure that participants were not primed to produce either of the lexical alternatives by a partner description. 

\subsection{Lexical Alternative Familiarisation Process}
So as to ensure that participants had knowledge of possible lexical alternatives, they were given time at the start of the experiment to familiarise themselves with the target images and the Hiberno and American-English terms for each of the 18 target pictures presented. Participants were informed they would be asked about the names of the picture after the task was completed so as to ensure they studied these in detail. 

\subsection{Interlocutor Conditions}
Participants interacted with either a human or computer partner (Partner Type) which used either a US accented or Irish accented voice (Partner Nationality), in a 2x2 between participants design. Similar to other studies \cite{branigan2011role,cowan2015does}, although participants were told that they were interacting with another partner, all item descriptions were in fact pre-recorded. For the US accented computer partner condition, descriptions of each item were recorded using Cereproc’s Cerevoice Hannah (US)\footnote{https://www.cereproc.com/en}.  For the Irish accented computer partner  condition, the same descriptions of each item were recorded using Cereproc’s Cerevoice Caitlin (Irish)\footnote{https://www.cereproc.com/en}. For the Irish and US accented human conditions, descriptions of each item were recorded by an Irish and US accented member of the research team. So as to ensure consistency in the interpretation of the conditions, participants were also  informed at the start of the experiment whether they were about to interact with another person (in the human partner conditions) or a computer (in the computer partner conditions). They were also told whether their interlocutor was Irish or American before commencing the game. 

\subsection{Procedure}
The research received ethical approval through the University's ethics procedures for low risk projects. Hiberno-English speakers, recruited from staff and students in a European University were recruited via email and were randomly allocated to conditions. Upon arrival they were welcomed by the experimenter, given information about the experiment and asked to give consent to take part in the study.  Following participants’ giving consent to take part, they were further briefed on the nature of the experiment task. They were told that they were to play a game with a partner, who was in another room. They were told that the aim of the game was to name selected pictures on the computer screen to their partner and pick out the ones that they name to them. Participants were encouraged to do this as quickly and as accurately as possible. Prior to starting the game, the researcher provided the list of Hiberno and American English terms for the 18 target items so participants could familiarise themselves with the possible lexical alternatives. During this period of familiarisation, the researcher informed participants they were stepping out of the room in order to get the partner ready for the task in a separate room. This was done in both the computer and human partner conditions to ensure that participants believed that they were in fact playing the game with another partner and not just the machine in front of them. Upon returning, the researcher informed participants who their partner was, based on one of the 4 interlocutor conditions. Both partners were described as being in another room and connected over a network.

Participants were then given further detailed instructions on the game. These instructions specified that participants should avoid describing the picture and were encouraged to use the names of the objects in the pictures. Participants were also given instructions on which buttons to press on the computer so as to progress through the turns. This was followed by a practice trial of the game involving 8 turns (4 matching and 4 naming). Once participants confirmed they fully understood the task, the full game was launched. Upon completion of the game, participants filled out an online demographics questionnaire. Here they provided details about: their age, sex and nationality; confirmed they were native or near native Hiberno-English speakers; levels of general and speech-based technological experience; how frequently they consumed US media and their existing knowledge of American English. They were also asked to comment on their experience of the experiment with particular reference to their partner and the game they played. Finally, participants were fully debriefed about the purposes of the study and thanked for taking part.

\section{RESULTS}
\subsection{Item effectiveness and lexical alternatives}

The first step in the analysis was to identify the effectiveness of the items developed for the game in eliciting the predicted American and Hiberno-English lexical alternatives. We found that out of the 18 items developed, 7 elicited the defined American and Hiberno-English lexical alternatives identified. Other items included were either given the wrong names by all participants (suggesting issues with the images developed for those items) or gave a large variety of alternative responses so as to question their interpretative validity. These were therefore excluded from the data. The preliminary item list and the number of American-English, Hiberno-English and utterances coded as Other are included in Table 1. Those marked with * were included in the analysis. 

Across the 238 lexical choices generated by participants for the final 7 target items, 40 (16.8\%) were American-English names, 187 (78.6\%) Hiberno-English names with 11 (4.6\%) being categorised as Other (coded as NAs for the analysis).

\subsection{Lexical alternatives analysis}

Because of the binary nature of the dependent variable (i.e the use of American or Hiberno-English lexical items) mixed effects logistic regression was run to analyse the data using the \emph{lme4} package (Version 1.1.19) \cite{bates2012package} in R (Version 3.5.2) \cite{rcore}. This analysis models the impact of the fixed effects in question (in this case, partner nationality and partner type) on the log odds of using an American English name in a target item description. These models allow us to take random effects due to item and participant variation into account in the analysis, increasing the statistical power \cite{singmann2017introduction} whilst also negating the need for separate partner and items analysis, common in experimental psycholinguistic research (see \cite{barr_random_2013, clark1973language}). The maximal model (see \cite{barr_random_2013}) did not converge so random effects were simplified until convergence was reached, whist ensuring that large correlations between random effects were minimised. The final model included by participants and by item random intercepts. The model syntax is displayed in Table 2. The outcome variable was releveled to ensure that the model refers to the log odds of producing American-English names in the fixed effects specified. The model estimates are shown in Table 2. 

\begin{figure}
  \includegraphics[width=0.4\textwidth]{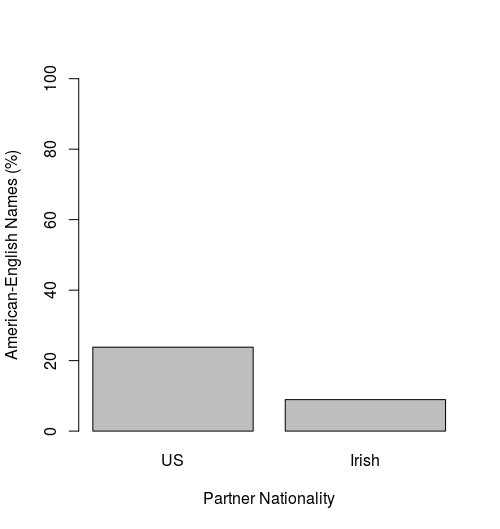}
  \caption{Total percentage of American-English names used by partner nationality}
  \label{fig:rplot}
\end{figure}

The results of the mixed effects analysis show that participants seemed less likely to produce American-English lexical alternatives when in the Irish partner conditions than in the US partner condition (Unstandardised $\beta$= -1.17, \emph{z}= 2.25, \emph{p}=.024) supporting H1. The number of lexical alternatives used are shown in Table 3. The total percentage of American-English lexical items used in the partner nationality conditions are shown in Figure \ref{fig:rplot}. 

This did not seem to vary by partner type. The frequency of American-English terms being used within the partner nationality conditions was not statistically different across partner types (Unstandardised $\beta$= -0.44,\emph{z}= 0.93, \emph{p}=.351), disconfirming H2.

\section{DISCUSSION}
Our study aimed to identify whether cues that may affect perceptions of linguistic knowledge, such as accent, may influence lexical choices in HMD. We found that participants were more likely to use American-English lexical items (and less likely to use Hiberno-English terms) when interacting with a US accented partner than with an Irish accented partner. Contrary to our hypothesis, the likelihood of using terms that corresponded with a particular accent was similar regardless of whether people were told they were talking to a machine or a human. Below we discuss the results and their implications for understanding the drivers of lexical adaptation in HMD.

\begin{table*}[t]
\caption{Summary of fixed and random effects on lexical alternatives-Logistic mixed effects model}  Model: Response\textasciitilde{}Partner Nationality*Partner Type+(1|Participant)+(1|Item) \\\hspace{\textwidth}
\begin{tabular}{lllll}
Fixed Effect                    & Unstandardised $\beta$ & SE $\beta$ & Wald-z & \textit{p}      \\ \hline
Intercept                       & -1.18                                   & 0.56                        & -2.13  & \textbf{.033 *} \\
Irish Partner                   & -1.17                                   & 0.52                        & -2.25  & \textbf{.024 *} \\
Computer Partner                & -0.44                                   & 0.47                        & -0.93  & .351            \\
Irish Partner* Computer Partner & -0.92                                   & 0.97                        & -0.95  & .341            \\ \hline
                                &                                         &                             &        &                 \\
Random Effects                  &                                         &                             &        &                 \\ \hline
Group                           & SD                                      &                             &        &                 \\ \hline
Participant                     & 0.154                                   &                             &        &                 \\
Item                            & 1.170                                   &                             &        &                
\end{tabular}
\end{table*}

\begin{table*}[t]
\caption{Lexical alternative frequencies by partner nationality and partner type}
\begin{tabular}{llllll}
Partner Nationality & Partner Type & N  & Hiberno-English Names (N, \%) & American-English Names (N,\%) & Other     \\ \hline
Irish               & Human        & 9  & 52 (82.5\%)                   & 8 (12.7\%)                    & 3 (4.6\%) \\
                    & Computer     & 7  & 44 (89.8\%)                   & 2 (4.1\%)                     & 3 (6.1\%) \\
                    & Total        & 16 & 96 (85.7\%)                   & 10 (8.9\%)                    & 6 (5.4\%) \\ \hline 
US                  & Human        & 9  & 45 (71.4\%)                   & 18 (28.6\%)                   & 0 (0\%)   \\
                    & Computer     & 9  & 46 (73.0\%)                   & 12 (19.0\%)                   & 5 (9.0\%) \\
                    & Total        & 18 & 91 (72.2\%)                   & 30 (23.8\%)                   & 4 (4.0\%)
\end{tabular}
\end{table*} 

\subsection{The impact of accent on partner models and lexical choice}
Our findings support the notion that design choices in HMD that cue differences in lexical knowledge (i.e. American or Hiberno-English lexicon) lead user's to make lexical choices with this perceived knowledge in mind. This has implications for speech interface design, highlighting the impact of synthesised voice choices, in particular system accent, on guiding user language use. The findings also give experimental support to the proposition that speech synthesis design impacts partner modelling and that partner models are significant drivers of HMD adaptation \cite{amalberti1993user,brennan1998grounding,LEBIGOT2007983,meddeb2010didn,branigan2011role}. It is likely that the synthesis accent in this case impacted user's perceptions of what lexicon the system favoured. The task in the experiment emphasised the choice of lexical alternatives that were related to the accents used, making this part of a users' model highly salient in interaction, influencing the choices made. That said, participants still predominantly used Hiberno-English terms in each condition. It seems though that partner accent does reduce this tendency. The influence of partner models in HMD may therefore not be absolute, but granular, with more egocentric production processes also at play in interaction. This echoes recent findings  on the role of other-centric and egocentric processes in dialogue \cite{duran2016toward,dale2018interacting}. 

\subsection{The dynamic nature of partner models in HMD}
Although not assessed here, the use of partner models over an interaction is likely to be highly dynamic. This will dependent on relevance of the models to decisions being made as well a the cognitive resources available for users to take these into account. Indeed the models themselves may also be dynamic, impacted by the interaction itself. As suggested in HHD research \cite{brennan2010two}, users will likely base their assumptions and perceptions on both a coarse-grained global model (e.g. assumptions of knowledge and abilities formed by stereotypes and expectations before interaction) and local experiences within the dialogue (e.g. feedback of comprehension via verbal and non verbal cues). Just as in HHD, local experiences in interaction may make a strong contribution to revising model assumptions, facilitating more partner specific model construction. Research has highlighted that partner models should be considered dynamic and adaptable over time \cite{fussell1989understanding,nickerson1999we}, yet little is known about how this occurs in HMD. Observing the dynamism of partner models across the course of a speech interface interaction, in particular how and when models are updated and across what specific dimensions, should be strongly considered in future research in this domain. 

\subsection{The role of partner type in user lexical choices}
Interestingly, we found no effects between partner types in our study. General approaches to partner modelling in HMD research would suggest ‘human’ and ‘computer’ operate as overarching global models that people compare when making sense of their experiences with speech interfaces \cite{leahu2013categories}. People tend to be acutely aware of the differences between humans and machines as dialogue partners, and consistently mention that they adapt their language accordingly \cite{cowan2017can,luger2016like,leahu2013categories}. A number of studies have also supported this quantitatively \cite{branigan2011role,KENNEDY198837}, yet recent work has shown this may not always be true \cite{cowan2015does,kiesler2008anthropomorphic}.  There may be a number of potential reasons for the lack of effect in our study. It may be that the simplicity of the task meant that participants did not see the difference between the partners' identity as salient. That is, recognition and comprehension of lexical terms to the level required for the game may have been seen as easy for either partner type. The humanness of synthesis may have also led people to use human communicative attributions as anchors for their partner models in the computer partner conditions throughout, without re-evaluation of their models. Recent work has emphasised the impact of humanness in design on potential models of system competence as a dialogue partner \cite{luger2016like, moore2017spoken, cowan2017can}. Again, because of the simplicity of the game and the apparent success of interaction with the system throughout, participants may have felt little need to re-evaluate these attributes and reflect on partner identity as important. Further work needs to disambiguate this, in particular observing the effects of experiences within dialogue on partner model adaptation. 

\subsection{Limitations}
Our research found that partner accent affected people's lexical choices, whereby they took partner's likely lexical knowledge into account when choosing words to use to describe objects. So as to make sure that participants perceived the accent and partner type consistently, we explicitly informed participants of the nationality and type of partner before they interacted in dialogue. This clarity and salience, although desirable for the consistency of the conditions in an experimental context, may not be apparent in more real-world interactions with accented systems. Indeed the type of dialogue interaction in this experiment is also constrained compared to more real-world interactions with systems. A referential communication task was used so as to observe lexical production systematically, whilst also allowing us to compare our findings across previous work on partner modelling in HMD and HHD (e.g. \cite{branigan2011role, cowan2015does}). It important for further research to attempt to replicate these effects in more real-world  and in less constrained dialogue contexts. 

Participants were also given an opportunity to familiarise themselves with the lexical alternatives and the relevant images before the study. This was so as to ensure that the effect was due to user choices around lexical alternatives and not because of limitations of participants' lexical knowledge. It also served to constrain participant word choices within the bounds of specific alternatives. It is important to make clear that these lexical items were not presented during the communication game, minimising the potential for direct priming effects. This also ensured participants had more freedom to choose between words they would more regularly use and those in the lexical alternatives presented when in dialogue. 

\section{CONCLUSION} 
Our research set out to identify how partner accented speech could affect user language choices in human-machine dialogue and compare this to human-human dialogue. We found that accent and partner nationality perceptions have a significant effect on people's lexical choices, encouraging them to produce names from the lexicon that may be more familiar to people with that accent. In this case, participants were more likely to produce American names for objects when interacting with partners with an American accent than when interacting with those with an Irish accent. This did not vary depending on whether the partner was perceived to be a person or a machine. From this it is clear that partner models, which are seen to be impactful in HHD, may also drive lexical choices in HMD where this is deemed appropriate. Crucially this adds much needed theory based insight into what may drive language choices in HMD. It seems that, like HHD, partner models may play a significant role in influencing language production in HMD. Speech interface design choices, such as which accents to use in speech synthesis, may not only influence user experience, but may also be critical in the mechanisms that guide user interaction behaviour.


\begin{acks}
This research was funded by the Irish Research Council COG-SIS Project (R17339). 
\end{acks}

\balance
\bibliographystyle{ACM-Reference-Format}
\bibliography{cui2019}

\end{document}